\documentclass[3p,times,twocolumn]{elsarticle}

\usepackage{ecrc}

\volume{00}

\firstpage{1}

\journalname{Nuclear and Particle Physics Proceedings}

\runauth{}

\jid{nppp}

\jnltitlelogo{Nuclear and Particle Physics Proceedings}

\usepackage{amssymb}

\usepackage[figuresright]{rotating}
\usepackage{subcaption}
\usepackage[utf8]{inputenc}
\usepackage{amsmath}

\begin{document}

\begin{frontmatter}



\dochead{}

\title{Rapidity-dependent jet energy loss in small systems with finite-size effects and running coupling}


\author[mc]{Chanwook Park}
\author[mc,bnl]{Chun Shen}
\author[mc]{Sangyong Jeon}
\author[mc]{Charles Gale}

\address[mc]{Department of Physics, McGIll University, 3600 University Street, Montreal, QC, H3A 2T8, Canada}
\address[bnl]{Physics Department, Brookhaven National Laboratory, Upton, NY 11973, USA}

\begin{abstract}
Longitudinal dynamics of particle production and rapidity-dependent jet energy loss are investigated in small and asymmetric colliding systems.
We utilize an improved version of \textsc{martini} in which two improvements are implemented to calculate the effect of the strongly coupled QGP droplet on jet energy loss.
We show that those realistic prescriptions improve the results  of nuclear modification factor calculations.
We also observe visible energy loss of jets in a thermal background of high-multiplicity p-Pb collisions, and a clear correlation between the energy loss and elliptic flow coefficients for energetic particles.
We conclude that systematic measurements of jet quenching in central collisions of small systems can support the formation of the QGP droplet.

\end{abstract}


\end{frontmatter}


\section{Introduction}
\label{sec:Intro}

In LHC and RHIC experiments, strong collective behavior is being observed in high multiplicity events in p-p and p-A collisions, suggesting that quark-gluon plasma can be created in such small systems~\cite{Chatrchyan:2013nka,Aad:2013fja,ABELEV:2013wsa}.
In this work, we introduce two improvements to the treatment of inelastic processes in \textsc{martini}~\cite{Schenke:2009gb}: finite formation time for emission, and the running coupling constant, which enable us to study jets in small QGP
droplets.
Those two features are expected to be essential in small systems~\cite{Shen:2016egw}.

Using this improved version of \textsc{martini}, we analyze the effect of the formation time and the running coupling in the particle production rate for a given rapidity regime, and calculate jet energy loss by the QGP droplet created in small and asymmetric systems.
We find sizeable medium-induced jet energy loss in high-multiplicity events of p-Pb collisions, in which temperatures can be comparable to those realized in heavy ion collisions.
Furthermore, we show that the elliptic flow coefficient for energetic particles in rapidity space are closely related to the amount of jet quenching.

\section{MARTINI: New developments}
\label{sec:New}

\begin{figure*}[t]
	\centering
	\begin{subfigure}[b]{0.36\textwidth}
		\centering
		\includegraphics[width=\textwidth]{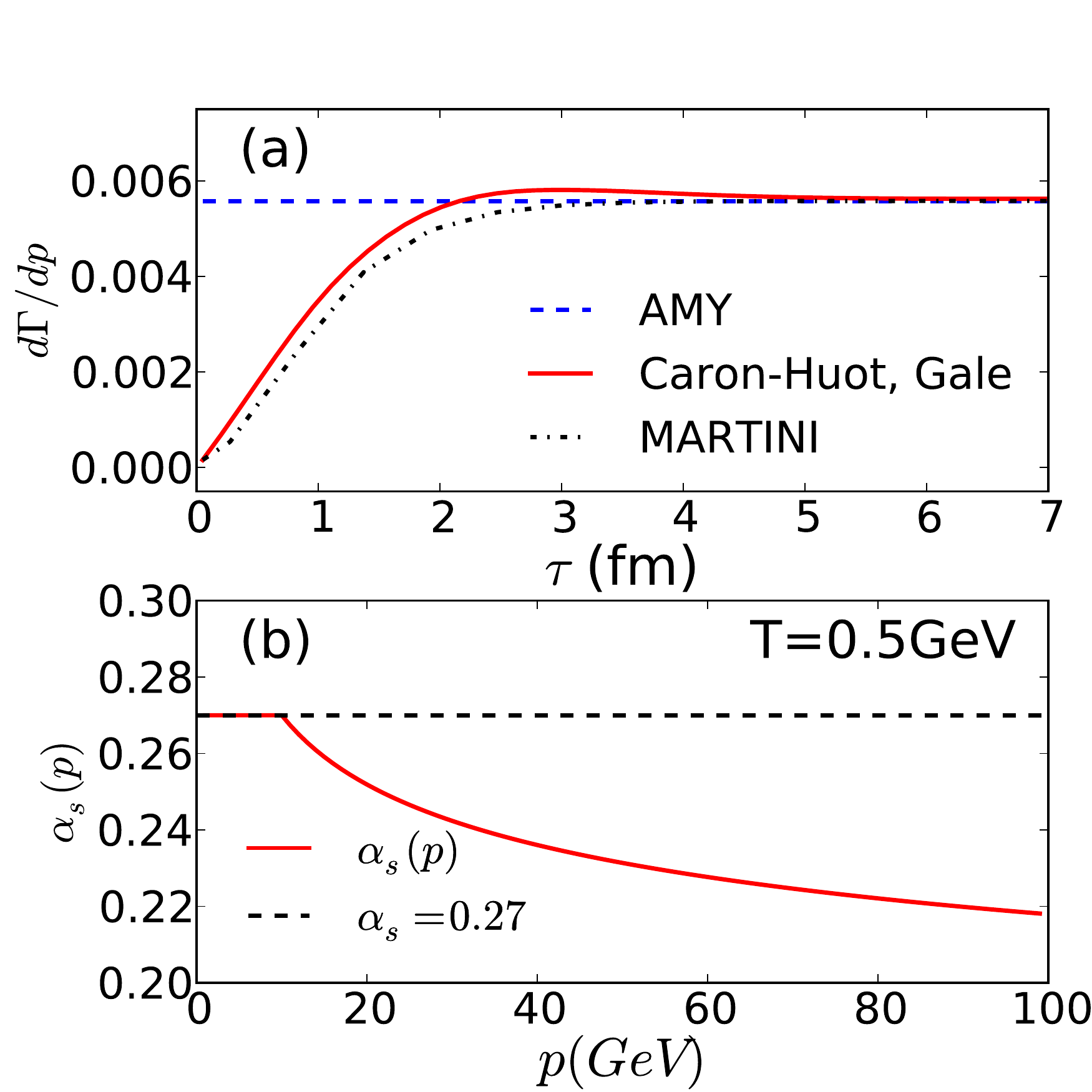}
	\end{subfigure}
	\begin{subfigure}[b]{0.3\textwidth}
		\centering
		\includegraphics[width=\textwidth]{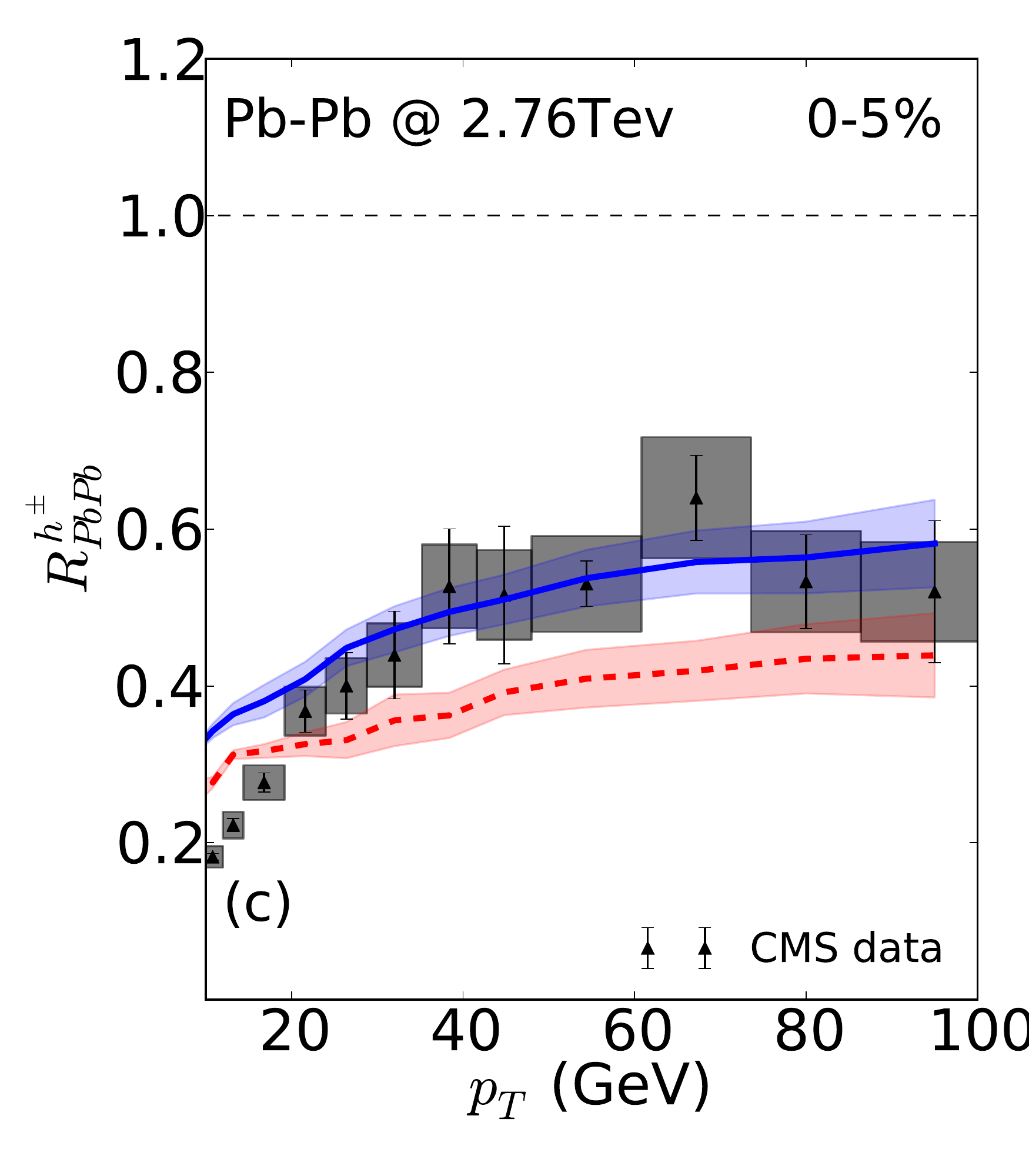}
	\end{subfigure}
	\begin{subfigure}[b]{0.3\textwidth}
		\centering
		\includegraphics[width=\textwidth]{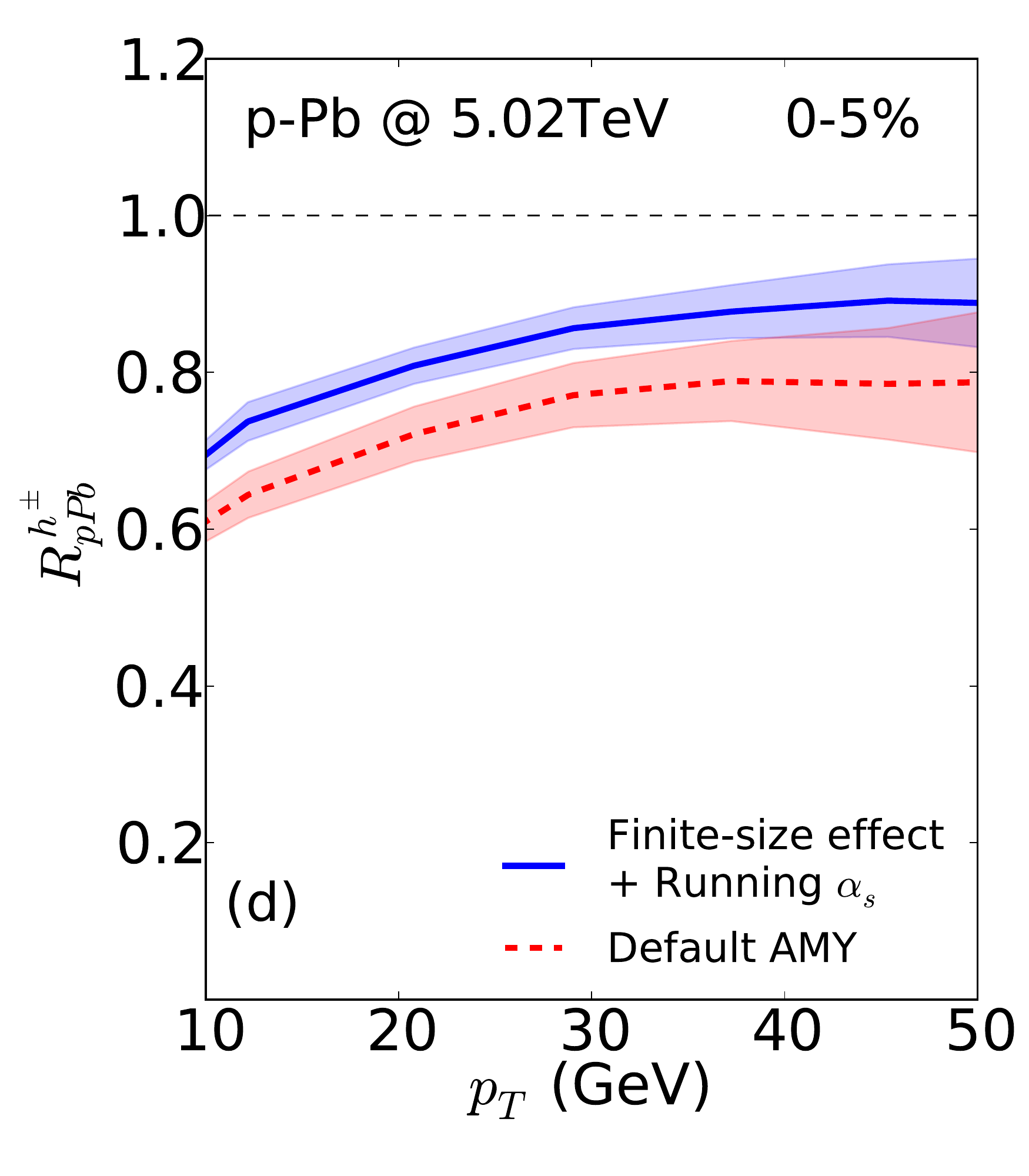}
	\end{subfigure}	\caption[]{(a): The rate for a 50 GeV quark emitting a 5 GeV gluon. (b): Running coupling constant, $\alpha_s$, as a function of momentum, $p$, of a particle at temperature, $T$= 0.5 GeV. (c): Charged hadron nuclear modification factor $R_{\rm AA}$ in 0-5\%  Pb-Pb collisions at $\sqrt{s}$ = 2.76TeV, compared with the ALICE measurement. (d): Charged hadron $R_{\rm pPb}$ in 0-5$\%$ p-Pb collisions at $\sqrt{s} =$ 5.02 TeV. In (c-d), the red curves correspond to the default AMY radiation rate, while the blue curves include the improved AMY rate described in the text.}
	\label{fig:the_two_modellngs}
\end{figure*}

In contrast to the soft scattering processes in a thermal medium, the formation time of the inelastic radiations increases with the energy of a parton.
The moment of the emission is not uniquely defined within the formation time.
Hence, during that time, a hard parton and an emitted parton are coherent and additional emissions from the two partons are prohibited until they are fully separated.
This interference between the two partons highly suppresses the radiation rates at early times after the original radiation.

In order to take the effects of finite formation time into account in the momentum-space AMY formalism~\cite{Arnold:2002zm}, the light-cone path integral formalism~\cite{Zakharov:1996fv} reformulated in Ref.~\cite{CaronHuot:2010bp} is written as
\begin{align} \label{eq:reformulated_equation}
	\frac{d\Gamma^a_{bc}(t)}{dk} & \equiv \frac{P^{a(0)}_{bc}(x)}{\pi p} 
	\times \textrm{Re} \int^t_0 dt_1 	\int_{\mathbf{q},\mathbf{p}} 
	\frac{i\mathbf{q}\cdot\mathbf{p}}{\delta E(\mathbf{q})}  \nonumber \\
	& \times \mathcal{C}(t)K(t,\mathbf{q};t_1,\mathbf{p}).
\end{align}
Here, $d\Gamma^a_{bc}(t)/dk$ is defined as the rate of the $a \to b + c$ radiation process and $P^{a(0)}_{bc}(x)$ is the leading order DGLAP splitting kernel.
The energy denominator $\delta E(\mathbf{p})$ is given by~\cite{CaronHuot:2010bp}
\begin{equation}
	\delta E(\mathbf{p}) = \frac{p\mathbf{p}^2}{2k(p-k)} + \frac{m^2_b}{2k} 
	+ \frac{m^2_c}{2(p-k)} - \frac{m^2_a}{2p},
\end{equation}
where $m_i$ is the thermal mass of the parton $i$.
The kernel $K(t,\mathbf{q};t_1,\mathbf{p})$ is the propagator associated with the light-cone Hamiltonian.
In momentum space, the time-dependent $\mathcal{C}(t)$ acts as the Boltzmann collisional operator~\cite{CaronHuot:2010bp}.
As noted in \cite{CaronHuot:2010bp}, the effects of the finite formation time on radiation rates become significant if we explore the higher energy regime.

To implement this quantum-mechanical phenomena in the Monte-Carlo event generator, we mimicked the rates calculated in \cite{CaronHuot:2010bp} in the following way.
After an emission, the two partons independently undergo multiple soft scatterings and momentum broadening.
Once their phase-space separation satisfies the uncertainty principle $\Delta r_\perp > 1/2\Delta p_\perp$, they become fully separated and are permitted to radiate again.
Fig.~\ref{fig:the_two_modellngs} (a) shows a rate for a 50 GeV quark emitting a 5 GeV gluon.
The slight enhancement over the AMY rate shown in the figure is due to interference effects, and it is not easy to reproduce it perfectly in a Monte-Carlo method.
However the difference between this implementation and the quantum-mechanical calculation is small.

The running coupling constant $\alpha_s(\mu)$ was applied for radiation processes~\cite{Young:2012dv}.
For the renormalization scale, we use the averaged momentum transfer $\sqrt{\langle p_\perp^2 \rangle}$ between the mother parton and the daughter parton estimated as follows.
The definition of jet transport coefficient $\hat{q}$ is given by
\begin{equation}\label{def:qhat}
	\hat{q} = \frac{\langle p^2_\perp \rangle}{t_f},
\end{equation}
where $t_f$ is the formation time of an emission.
Combining Eq.~\ref{def:qhat} with $t_f = p / \langle p^2_\perp \rangle$, one gets 

\begin{equation}
	\sqrt{\langle p_\perp^2 \rangle} = (\hat{q}p)^{1/4}.
\end{equation}
Running coupling $\alpha_s(p)$ as a function of momentum $p$ is shown in Fig.~\ref{fig:the_two_modellngs} (b). 
A 20\% reduction in the strength of interaction at $p = 100$ GeV is found.

Both of these effects -- the finite-size effects and the effect of running coupling -- lead to a decrease in the energy loss rates induced by inelastic processes as shown in Fig.~\ref{fig:the_two_modellngs} (c-d).
As can be seen in the figure, this leads to overall a better description of the Pb-Pb collisions and predicts
$R_{\rm pA}$ to be around 0.8 -- 0.9.

\section{Rapidity-dependent jet energy loss}
\label{sec:rapidity_jet_eloss}

\begin{figure*}[t]
	\centering
	\begin{subfigure}[b]{0.36\textwidth}
		\centering
		\includegraphics[width=\textwidth]{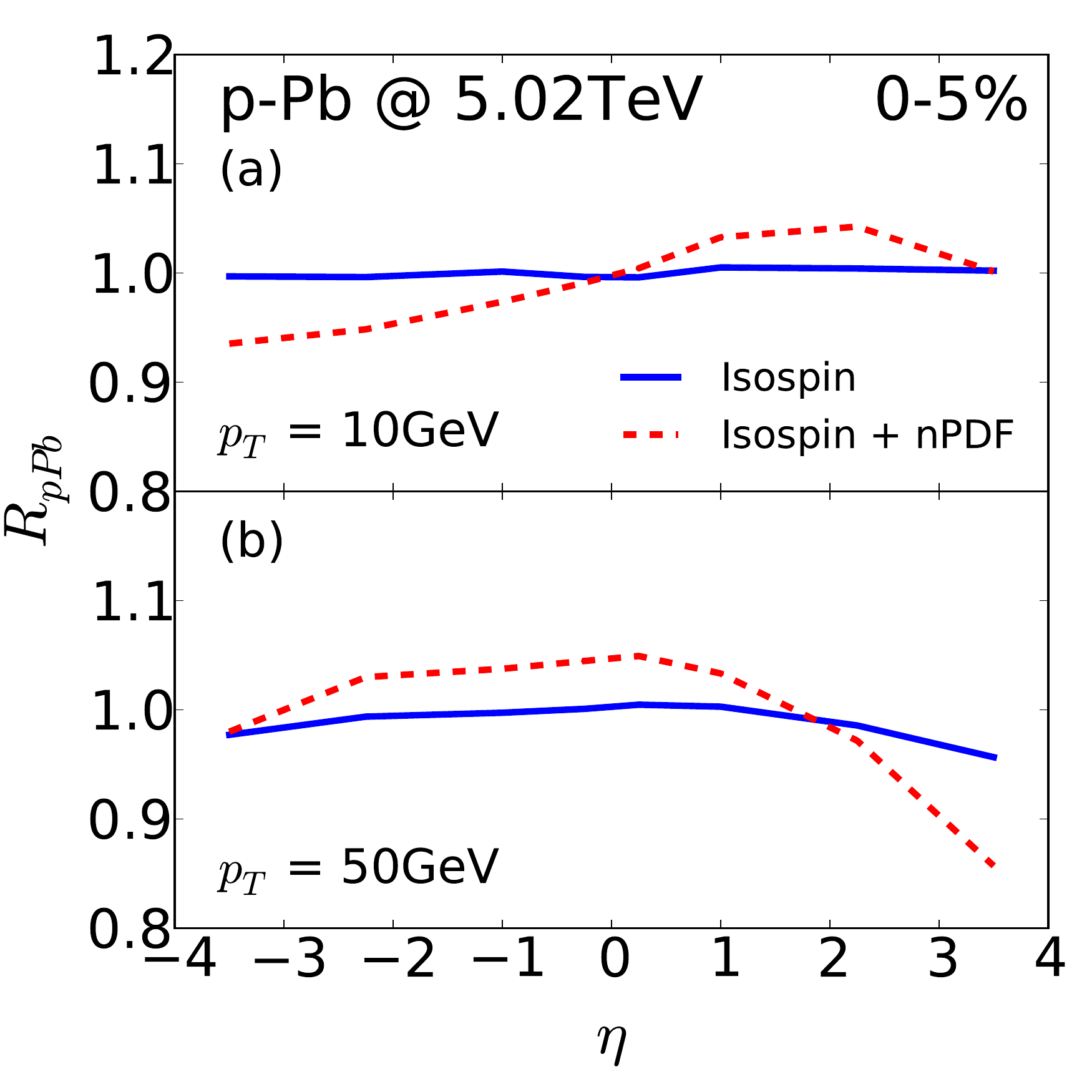}
	\end{subfigure}
	\begin{subfigure}[b]{0.58\textwidth}
		\centering
		\includegraphics[width=\textwidth]{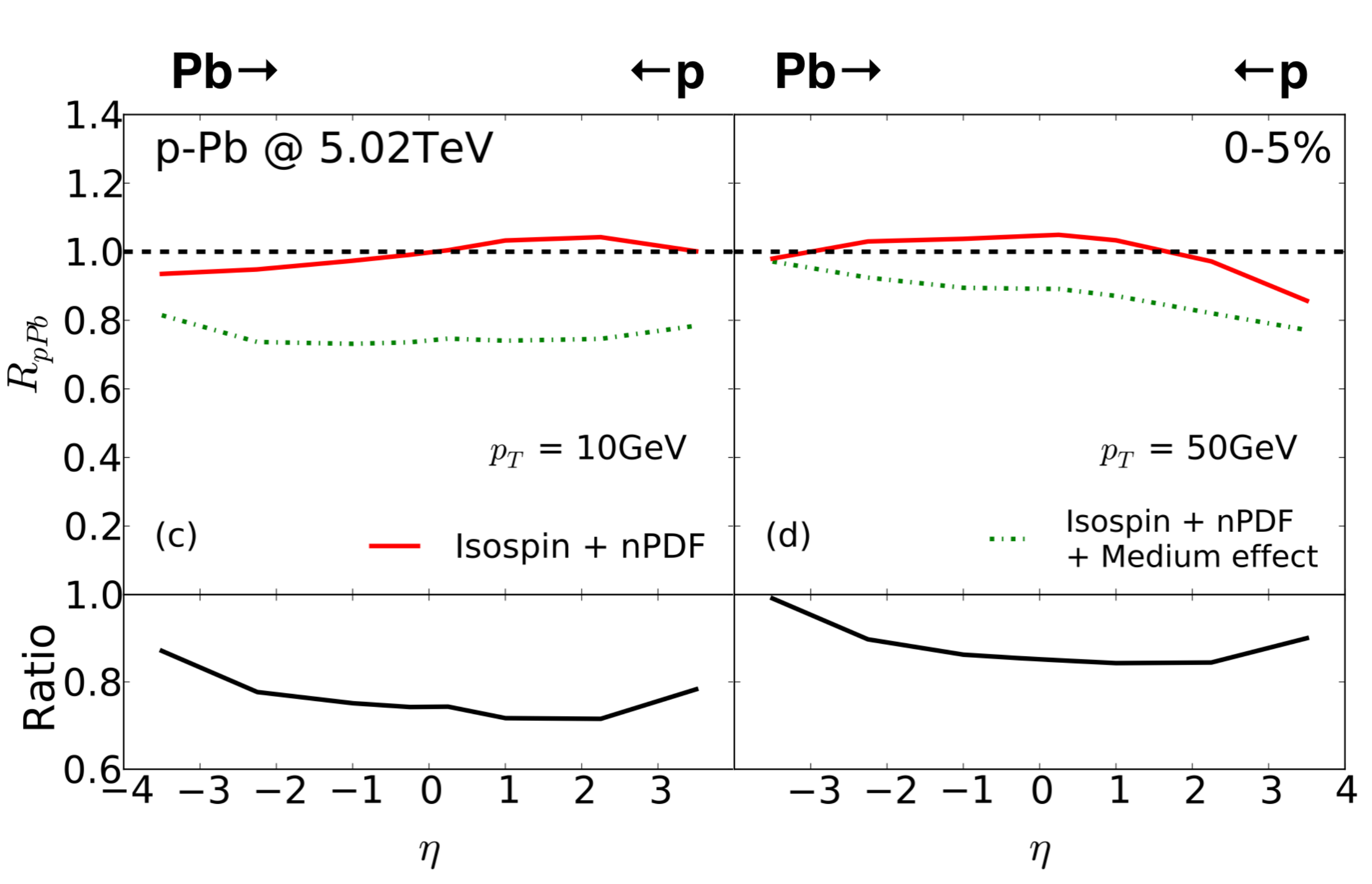}
	\end{subfigure}
	\caption[]{Nuclear modification factor $R_{\rm pPb}$ at 10 GeV and 50 GeV as a function of rapidity in 0-5$\%$ p-Pb collisions at $\sqrt{s}$ = 5.02 TeV. Left panel: the isospin effects (red) and the isospin + the nuclear PDF (blue). Right panel: additional medium-induced jet energy loss effect on top of the isospin effect and the nuclear PDF (green). Note the red curves in the two panels are identical. The black curves correspond to the ratio between the red and the green curve.}
	\label{fig:Eloss_effects}
\end{figure*}

Colliding asymmetric systems such as p-Pb offer an opportunity to study underlying physics that governs different rapidity regimes.
To model the non-trivial longitudinal dynamics of such collisions, we used the thermal media created by \textsc{music}~\cite{Schenke:2010nt}, which allows full 3+1 dimensional hydrodynamic calculations~\cite{Shen:2016zpp}.
Medium fluctuations were taken into account by using event-by-event simulations with Monte-Carlo Glauber initial conditions. 
Centrality classes were determined by the initial entropy density $dS/dy$, which is highly correlated with the final state charged-hadron multiplicity $dN^{ch}/d\eta$~\cite{Shen:2015qta}.
We simulated jet evolution in the hydrodynamic background using the improved \textsc{martini} at $\sqrt{s}$ = 5.02 TeV.
Parton sampling was done in accordance with CTEQ6L~\cite{Pumplin:2002vw} parton distribution function (PDF) and EPS09LP~\cite{Eskola:2009uj} nuclear PDF.

Fig.~\ref{fig:Eloss_effects} shows $R_{\rm pPb}$ as a function of rapidity in the 0-5$\%$ centrality bin at (a) 10 GeV and (b) 50 GeV.
In our setup, lead (proton) is assigned to the positive (negative) rapidity.
The blue curves only include the isospin effect, the difference of the $u$ and $d$ quark PDFs, while the isospin effect and the nuclear PDF for Pb nuclei are included in the red curves.
As low energy particles address low $x$ values, for which the PDFs of $u$ and $d$ quarks are almost identical~\cite{Pumplin:2002vw}, the blue curve in Fig.~\ref{fig:Eloss_effects} (a) is essentially unity.
On the other hand, an enhancement is observed in the forward rapidity regime in the curve with nuclear effects. There, $x$ from Pb is large enough that anti-shadowing effects appears~\cite{Eskola:2009uj}.
The shadowing effect is also observed in the backward side.
For higher energy particles, the difference between $u$ and $d$ quark PDFs becomes visible, and a slight decrease in $R_{\rm pPb}$ from unity is observed in the very forward and backward regimes.
This deviation would be larger as the energy of particle increases.
As the energy contribution from Pb increases, we find the anti-shadowing effect in a broad range of rapidity and a large depletion effect in the very forward rapidity region.

The energy loss effect generated by the evolving medium is shown in Fig.~\ref{fig:Eloss_effects} (c-d).
The green curves contain the additional medium effect on top of the isospin effect and the nuclear PDF, while in the red curve the medium effect is absent.
Visible jet energy loss is observed and the ratio of the two curves indicates that its effect is non-trivial in the rapidity space.
We find that the net medium-induced energy loss is closely related to the energy of particles and temperature of the background medium.
The ratios are also consistent with the charged-hadron multiplicity $dN^{ch}/dy$ in the final state~\cite{Shen:2016zpp}.
If observed in experiments, this can be a clear signature of QGP droplets created in central collisions of small systems.

\begin{figure}[t]
	\centering
	\centering
	\includegraphics[scale=0.4]{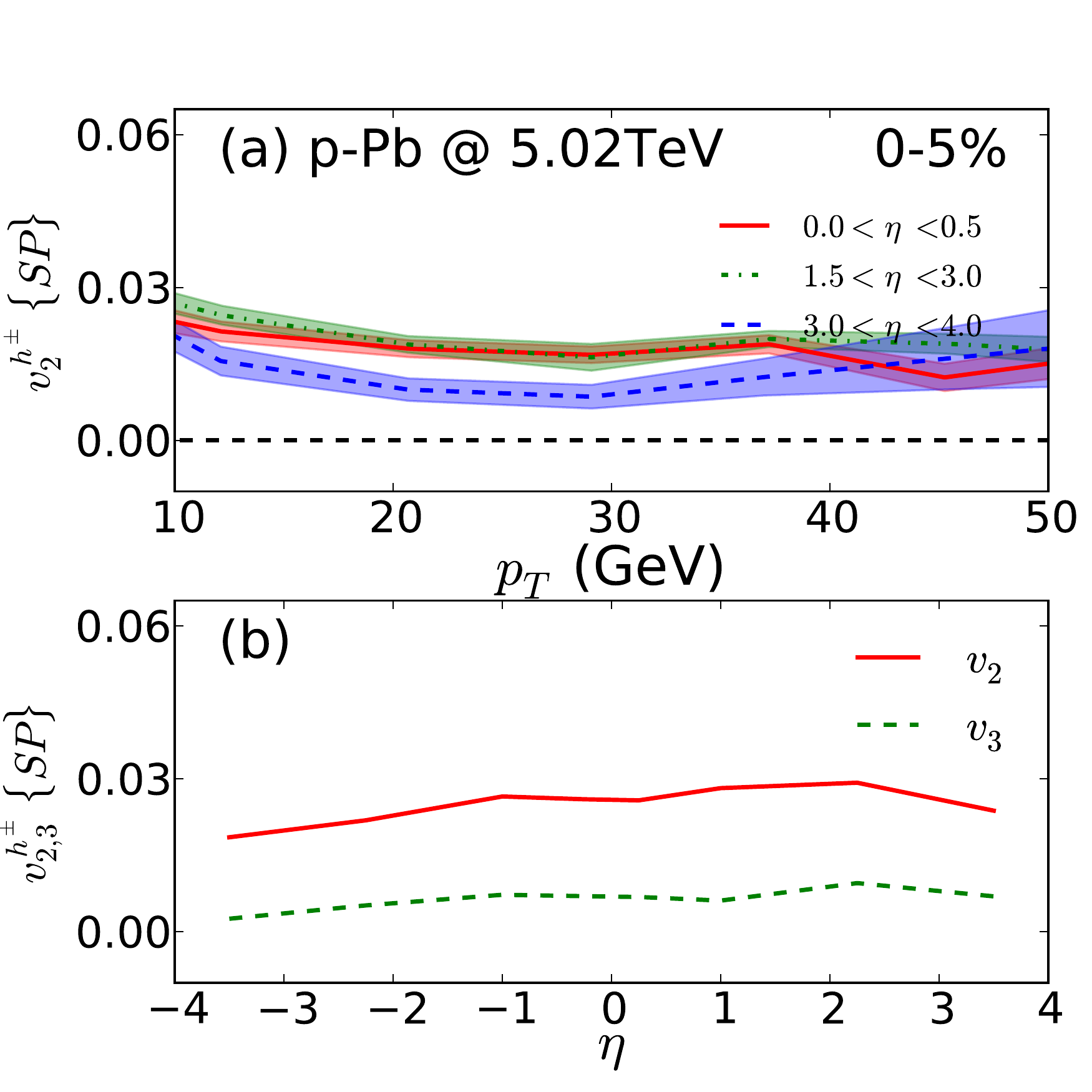}
	\caption[]{Elliptic flow coefficient $v_2$\{SP\} for energetic charged hadrons in 0-5$\%$ p-Pb collisions at $\sqrt{s}$ = 5.02 TeV. (a): $p_T$ differential $v_2$\{SP\} in 3 different rapidity intervals. (b): $p_T$ integrated $v_{2,3}$\{SP\} from 10 to 50 GeV.}
	\label{fig:vn}
\end{figure}

The differential and $p_T$ integrated harmonic flow coefficients $v_{2,3}$ for energetic charged hadrons computed using the scalar product method~\cite{Luzum:2012da,Noronha-Hostler:2016eow} are shown in Fig.~\ref{fig:vn}.
Even for high energy particles, $10 < p_T < 50$ GeV, the non-zero values of the charged elliptic flows are observed.
Depending on the rapidity intervals, $v_2$\{SP\} shows 1-3$\%$ and exhibits non-trivial rapidity dependence.
In Fig.~\ref{fig:vn} (b), we also find that, compared with the ratios shown in Fig.~\ref{fig:Eloss_effects} (c-d), the strength of the harmonic flows for high-energy particles is correlated with the medium-induced energy loss, e.g.\ large medium-induced energy loss indicates a large flow coefficient.

\section{Conclusion}
\label{sec:con}

In this work, we introduced two realistic energy loss models into \textsc{martini}, and demonstrated that the both of those models play an essential role in describing jet quenching in Pb-Pb collisions.
Our study on the particle production rate in p-Pb collisions together with the anisotropic flow calculations for energetic particles predicted sizeable energy loss, which indicates the formation of QGP droplets in such small collision systems.
More in-depth analyses on the medium structure and jet energy distribution are in progress.

\section{Acknowledgements}

This work was supported in part by the U.S. Department of Energy, Office of Science under contract No. DE- SC0012704 and the Natural Sciences and Engineering Research Council of Canada. C.S. gratefully acknowledges a Goldhaber Distinguished Fellowship from Brookhaven Science Associates, and C. G. gratefully acknowledges support from the Canada Council for the Arts through its Killam Research Fellowship program.
Computations were made in part on the supercomputer Guillimin from McGill University, managed by Calcul Qu\'ebec and Compute Canada.
The operation of this supercomputer is funded by the Canada Foundation for Innovation (CFI), NanoQu\'ebec, RMGA and the Fonds de recherche du Qu\'ebec - Nature et technologies (FRQ-NT).




\nocite{*}
\bibliographystyle{elsarticle-num}
\bibliography{Park_C}







\end{document}